\documentclass[preprint]{revtex4}

\usepackage{epsfig}                 
 
\begin{document}
\draft
\title{A Position-Space Renormalization-Group Approach for Driven Diffusive Systems
Applied to the Asymmetric Exclusion Model}
\author{Ivan T.\ Georgiev and Susan R.\ McKay}
\affiliation{University of Maine, Orono, ME 04468}
\date{\today}
\begin{abstract}
  
This paper introduces a position-space renormalization-group approach for nonequilibrium systems and applies the method
to a driven stochastic one-dimensional gas with open boundaries. The dynamics are characterized by
three parameters:  the probability $\alpha$ that a particle will flow into the chain to
the leftmost site, the probability $\beta$ that a particle will flow out from the rightmost
site, and the probability $p$ that a particle will jump to the right if the site to the right is
empty.  The renormalization-group procedure is conducted within the space of these transition
probabilities, which are relevant to the system's dynamics. The method yields a critical point at $\alpha_c=\beta_c=1/2$,
in agreement with the exact values, and the critical exponent $\nu=2.71$, as compared with the exact value  $\nu=2.00$.

\end{abstract}

\maketitle

 \section{INTRODUCTION}
Driven diffusive systems exhibit a variety of nonequilibrium phase transitions between steady states of the system (See, for example, \cite{ziabook}).  We present a general position-space renormalization-group approach for these systems and illustrate its application to the asymmetric exclusion model  \cite{Der93, Der_book, Krug}.  This model, a one-dimensional lattice gas with open boundary conditions, provides an excellent testing ground for new methods, since it exhibits both first and second order phase transitions, and the exact solution is known \cite{ren1, ren2}. We apply the method to a system with stochastic dynamics, one in which a master equation describes its evolution.  Recursion relations link the model parameters that are relevant for the system's dynamics and criticality on various length scales.  In the asymmetric exclusion model, these parameters are expressible in terms of one and two site probability distribution functions.  In order to construct the recursion relations, we exploit the exact general form of the solution \cite{Der92} and the knowledge of the steady state current in each different region.  When the solution is not known, the functional dependence of the current on model parameters can be obtained, at least approximately, from the mean field solution.  For the asymmetric exclusion model, a mean field treatment yields the exact current as a function of $\alpha$ and $\beta$, the probability rates for particles entering and leaving the system respectively.
There have been several recent papers on position-space renormalization for reaction-diffusion systems ~\cite{RG_RDS1, RG_RDS2, RG_RDS3} that successfully study these models. They use the fact that these models can be related to the ground state of a suitably defined quantum hamiltonian and then use the methods available for quantum spin systems. Another recent work ~\cite{RG_RDS4} investigates mainly the asymmetric exclusion model, as we do in this paper, by developing a position-space rescaling procedure that preserves the density and the current in the chain and calculating the dynamical critical exponent of the model.

The model consists of a one-dimensional open chain of $N$ sites. Each site {\it i} can be occupied ($\tau_i=1$)
or empty ($\tau_i=0$). A particle can hop to its right neighbor provided that the neighboring site is empty. The dynamics are sequential: at each time step {\it dt}, we choose at random a pair of sites $(i,i+1)$ and, if site $i$ is occupied
and site $i+1$ is empty, then the particle at the $i^{th}$ site will jump to the right with probability $dt$:
\begin{eqnarray}
\tau_i(t+dt)&=&1, {\rm \; with\; probability\;} x_i=\tau_i(t)+[\tau_{i-1}(t)(1-\tau_i(t))-
\tau_i(t)(1-\tau_{i+1}(t))]dt \nonumber\\
\tau_i(t+dt)&=&0, {\rm \; with\; probability\;} 1-x_i, {\rm \; where \;} \; i \;  \in \{1,...,N-1\}.
\label{EQ1}
\end{eqnarray}
All of the other sites do not change. The boundary sites are treated in the following way: when
 the chosen pair is $(0,1)$, where site $0$ represents the left source of particles, a particle is injected
into the chain with probability $\alpha dt$ if the first site of the chain is empty:
\begin{eqnarray}
\tau_1(t+dt) &=& 1,\; \text{with probability $x_0=\tau_1(t)+\alpha [1-\tau_1(t)] dt$}\nonumber\\
\tau_1(t+dt) &=& 0, \; \text{with probability $1-x_0$}.
\label{EQ2}
\end{eqnarray}

When the chosen pair is $(N,N+1)$, where the $N+1$ site represents the right boundary of the chain, the
particle at site $N$, if it is occupied, will flow out of the chain with probability $\beta dt$:
\begin{eqnarray}
\tau_N(t+dt) &=& 1, {\rm \; with\; probability\;} x_N= (1- \beta) \tau_N(t)dt\nonumber\\
\tau_N(t+dt) &=& 0, {\rm \; with\; prob\begin{scriptsize}  \end{scriptsize}ability\;} 1-x_N.
\label{EQ3}
\end{eqnarray}

More general models have been studied \cite{Schutz, Sandow} introducing possibilities for the particles to
jump to a left neighbor and allowing particles at the left boundary to flow out of the chain
and particles at the right boundary to flow into the chain.

Now we want to investigate the appearance of the steady state distributions.  Averaging the above equations over the
events that may occur in one time step $dt$ and over the histories up to time $t$, one obtains \cite{Der_book}:
\begin{eqnarray}
\frac{d}{dt}\langle\tau_i\rangle &=& \langle\tau_{i-1}(1-\tau_i)\rangle-\langle\tau_{i}(1-\tau_{i+1})\rangle ,  {\rm \; where \;} \; i \;  \in \{1,...,N-1\}.\nonumber\\
\frac{d}{dt}\langle\tau_1\rangle &=& \alpha\langle 1-\tau_1\rangle-\langle\tau_1(1-\tau_2)\rangle\nonumber\\
\frac{d}{dt}\langle\tau_N\rangle &=& \langle\tau_{N-1}(1-\tau_N)\rangle-\beta\langle\tau_N\rangle
\label{EQ4}
\end{eqnarray}
Eqs.\ (\ref{EQ4}) serve as our basic equations for applying the renormalization-group procedure. The steady state of
the model is given in terms of $P_N(\tau_1,\tau_2,...,\tau_N)$, which are the probabilities of finding the specific
configuration represented by the occupation numbers $(\tau_1,\tau_2,...,\tau_N)$ in a chain with $N$ sites. In the long time
limit, the system reaches a steady state where the probabilities $P_N(\tau_1,\tau_2,...,\tau_N)$ do not change with time, i.e.

$$\frac{d}{dt}P_N(\tau_1,\tau_2,...,\tau_N) = 0,$$
\begin{equation}
\frac{d}{dt}\langle\tau_i\rangle = 0, \;etc.
\label{EQ5}
\end{equation}
The steady state solution can be obtained from the expression \cite {Der93, Der92}:
\begin{eqnarray}
P_N(\tau_1,\tau_2,...,\tau_N) &=& f_N(\tau_1,\tau_2,...,\tau_N)/Z_N, {\rm\;with\;}\nonumber\\
Z_N &=& \sum_{\tau_1=0}^1\sum_{\tau_2=0}^1...\sum_{\tau_N=0}^1f_N(\tau_1,\tau_2,...,\tau_N), {\rm\;and\;}\nonumber\\
f_N(\tau_1,\tau_2,...,\tau_N) &=& \langle W\vert\prod^N_{i=0}(\tau_iD+(1-\tau_i)E\vert V\rangle,
\label{EQ6}
\end{eqnarray}
where D and E are square matrices, $\langle W\vert$ and $\vert V\rangle$ are vectors satisfying:
\begin{eqnarray}
DE &=& D+E\nonumber\\
D\vert V\rangle &=& \frac{1}{\beta}\vert V\rangle\nonumber\\
\langle W\vert E &=& \frac{1}{\alpha}\langle W\vert .
\label{EQ7}
\end{eqnarray}

\begin{figure}[htb]
\epsfig{file=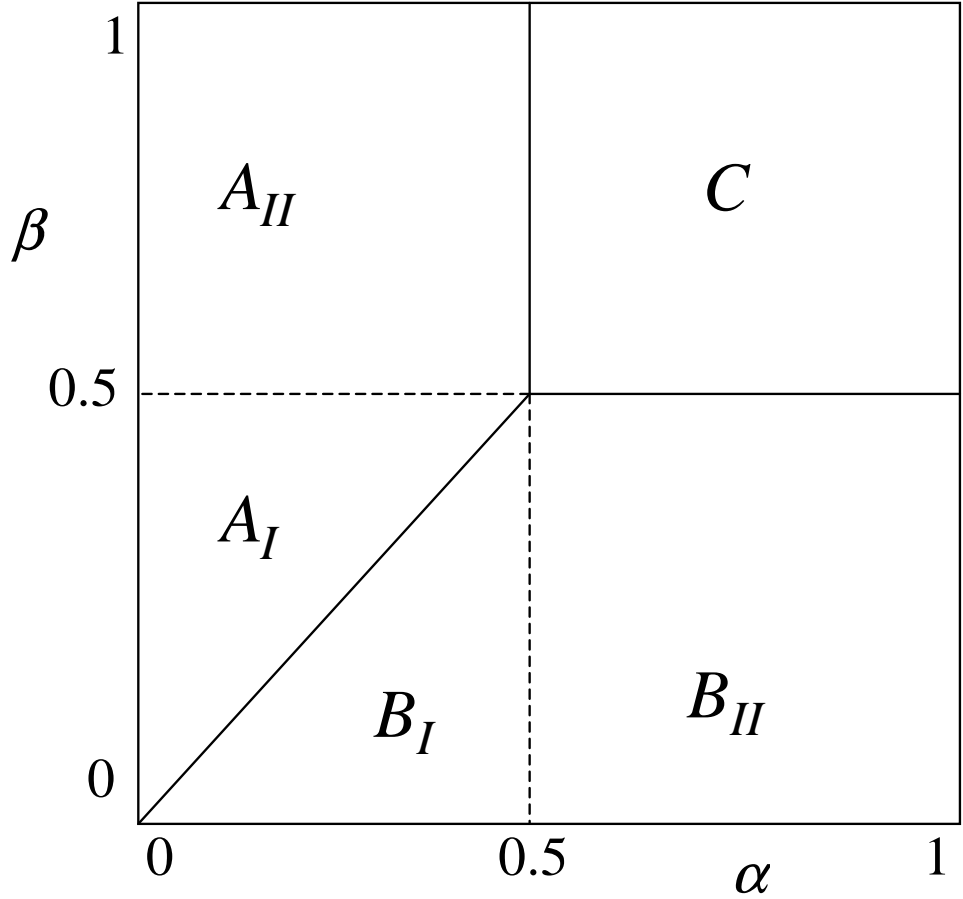, width=5in}
\caption{The phase diagram of the model. (See, for example, \cite{Schutz}.) The low density phase $A$ is divided into two phases $A_I$ and $A_{II}$, and
the high density phase $B$ into $B_I$ and $B_{II}$. The maximum current phase is labeled $C$. The bulk density in
the regions $A$, $B$, and $C$ is respectively: $\alpha$, $1-\beta$, and $1/2$. The lines $\alpha=0.5$ and $\beta = 0.5$
indicate second order phase transitions. The line $\alpha = \beta < 0.5$ is a first order phase transition. }
\label{fig1}
\end{figure}

As shown in Fig.\ \ref{fig1}, the system can be
in three phases: a low density phase $(A)$, a high density phase $(B)$, and the maximum current phase $(C)$.
The high and low density phases are separated by a first order phase boundary, and both are separated from the maximum
current phase by a second order phase boundary.
The steady state current is given by the formula \cite{Der93}:
\begin{equation}
J_N=\frac{\langle W\vert C^{N-1}\vert V\rangle}{\langle W\vert C^N\vert V\rangle},\;{\rm\;where\;}C \equiv D+E.
\label{EQ8}
\end{equation}
The current is a continuous function over the whole parameter space and, in the thermodynamic limit as $N\rightarrow \infty $,
has the simple form:
\begin{eqnarray}
J &=& \alpha (1-\alpha),\;{\rm\;for\;}\alpha <1/2{\rm\;and\;}\beta >\alpha\;\; (phase\;B)\nonumber\\
J &=& \beta (1-\beta),\;{\rm\;for\;}\beta <1/2{\rm\;and\;}\alpha >\beta\;\; (phase\;A)\nonumber\\
J &=& 1/4,\;\;\; \;\;\;{\rm\;for\;}\beta \geq1/2{\rm\;and\;}\alpha \geq 1/2\;\; (phase\;C).
\label{EQ9}
\end{eqnarray}
In applying the rescaling scheme we need only the above functional relation for the steady
state current which, for this system, can be obtained directly from a mean field treatment \cite{Der92}.  Thus the method 
is applicable to other types of systems, in which the exact solution is not known.

\section{MAIN RESULTS}
Here we illustrate the general rescaling procedure with a length rescaling factor of three.  Using a larger length rescaling 
factor would be expected to yield more accurate results, but also leads to substantially more involved algebra in the recursion relations.
The set $\{ \tau_1, \tau_2,...,\tau_N \}$ maps into the set $\{T_1, T_2, ..., T_{\tilde N} \}$, where we have used the majority rule to determine the state (empty or occupied) of the coarse-grained site and $\tilde N = N/3$. We assume that the matrix algebra remains the same after the blocking procedure, prohibiting any expansion of the parameter space. The rates $\alpha$ and $\beta$ are allowed to evolve under rescaling, while the rate for the forward jump $p=1$ is held constant.
From Eq.\ (\ref{EQ4}) it follows that:
\begin{eqnarray}
\alpha &=& \frac{\langle \tau_1(1-\tau_2)\rangle}{\langle 1-\tau_1\rangle}\nonumber\\
\beta &=& \frac{\langle \tau_{N-1}(1-\tau_N)\rangle}{\langle \tau_N\rangle}.
\label{EQ10}
\end{eqnarray}
Therefore for the rescaled parameters $\tilde{\alpha}$ and $\tilde{\beta}$ we have:
\begin{eqnarray}
\tilde{\alpha} &=& \frac{\langle T_1(1-T_2)\rangle}{\langle 1-T_1\rangle}\nonumber\\
\tilde{\beta} &=& \frac{\langle T_{N-1}(1-T_N)\rangle}{\langle T_N\rangle}.
\label{EQ11}
\end{eqnarray}
The one-site and two-site probability distributions in the coarse-grained chain can be expressed in
terms of three-site and six-site probability distribution functions of the original chain as follows:
\begin{eqnarray}
\langle 1C^{\tilde{N}-1}\rangle _T = \langle 111C^{N-3}\rangle _\tau + \langle 011C^{N-3}\rangle _\tau+\langle 101C^{N-3}\rangle _\tau+\langle 110C^{N-3}\rangle _\tau\nonumber\\
\langle 10C^{\tilde{N}-2}\rangle _T=\langle 111000C^{N-6}\rangle _\tau+\langle 111001C^{N-6}\rangle _\tau+\langle 111010C^{N-6}\rangle _\tau+...,\;etc.
\label{EQ12}
\end{eqnarray}
The sub-indexes $T$ and $\tau$ indicate the system on which the average is calculated. Working out each of these 
distributions, using the algebra represented by Eq.\ (\ref{EQ7}), we find that the rescaled values depend on the 
expression:
\begin{equation}
\frac{\langle W\vert C^{N-a}\vert V\rangle}{\langle W\vert C^N\vert V\rangle}=\frac{\langle W\vert C^{N-1}\vert V\rangle}{\langle W\vert C^N\vert V\rangle}\frac{\langle W\vert C^{N-2}\vert V\rangle}{\langle W\vert C^{N-1}\vert V\rangle}\;...\;\frac{\langle W\vert C^{N-a}\vert V\rangle}{\langle W\vert C^{N-a+1}\vert V\rangle},
\label{EQ13}
\end{equation}
where $a$ is a natural number. In the thermodynamic limit, each of these ratios becomes the current $J$, and this expression can be rewritten as:
\begin{equation}
\lim_{N\to \infty}\frac{\langle W\vert C^{N-a}\vert V\rangle}{\langle W\vert C^N\vert V\rangle}=J^a
\label{EQ14}
\end{equation}

\begin{figure}[htb]
\epsfig{file=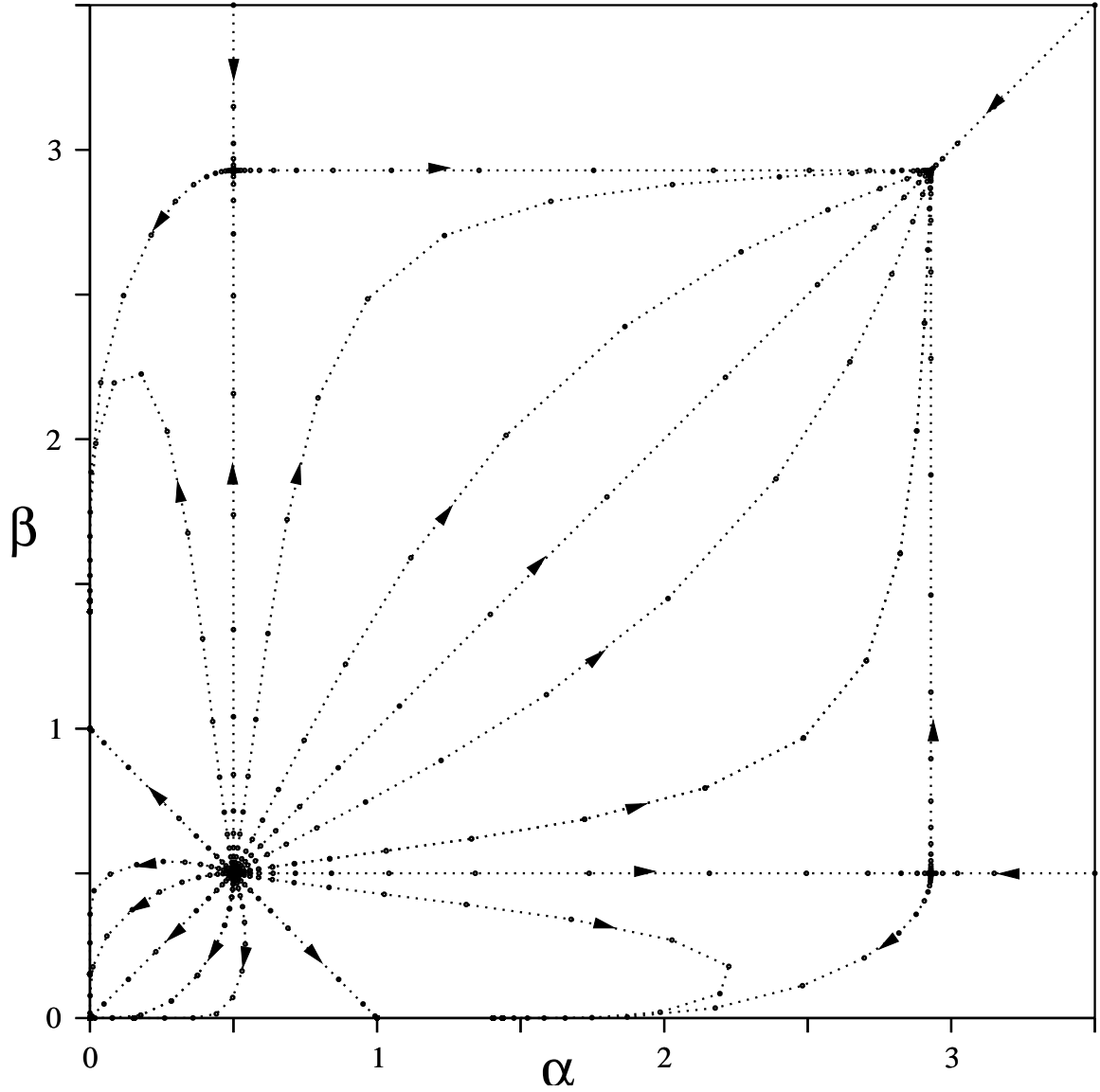, width=6in, height=6in}
\caption{Flow diagram for the totally asymmetric exclusion model.  Points start from the vicinity of the repulsive fixed point 
$\alpha_c=\beta_c=0.5$.  There are fixed points at: $(0,0),\; (0,1),\; (1,0),\; (0.5,2.929),\;(2.929,0.5)$ and  $(2.929, 2.929)$. Comparison with Fig.\ \ref{fig1} shows that the flow lines capture exactly the position of the first and second order phase transitions as well as the critical point. }
\label{fig2}
\end{figure}

After some algebra (detailed in the Appendix), one obtains from Eq.\ (\ref{EQ11}) using Eqs.\ (\ref{EQ7}), (\ref{EQ8}), (\ref{EQ12}), and (\ref{EQ14}) the final recursion relations:
\begin{eqnarray}
\tilde \alpha=\frac{ 3+4J+(6-\frac{3}{\alpha}-\frac{3}{\alpha^2})J^2+(4-\frac{1}{\alpha}+\frac{3}{\alpha^2}+\frac{3}{\alpha^3})J^3+(\frac{4}{\alpha^2}+\frac{3}{\alpha^3})J^4}{1+\frac{1}{\alpha}+\frac{1}{\alpha^2}+\frac{J}{\alpha^2}}\nonumber\\
\tilde \beta=\frac{ 3+4J+(6-\frac{3}{\beta}-\frac{3}{\beta^2})J^2+(4-\frac{1}{\beta}+\frac{3}{\beta^2}+\frac{3}{\beta^3})J^3+(\frac{4}{\beta^2}+\frac{3}{\beta^3})J^4}{1+\frac{1}{\beta}+\frac{1}{\beta^2}+\frac{J}{\beta^2}}
\label{EQ15}
\end{eqnarray}
From Eq.\ (\ref{EQ15}) the flow diagram displayed in Fig.\ \ref{fig2} is generated as follows. First we select initial values of $\alpha$ and $\beta$. These values determine the current $J$, which has different values in the $\alpha-\beta$ plane as specified by Eq.\ (\ref{EQ9}). Plugging the values of $\alpha$, $\beta$, and $J$, into Eq.\ (\ref{EQ15}) yields the rescaled values $\tilde \alpha$ and $\tilde \beta$. The rescaled current is obtained again from Eq.\ (\ref{EQ9}), using the rescaled values of $\alpha$ and $\beta$. This process is done iteratively to generate the full flow diagram.

The flow diagram shown in Fig.\ \ref{fig2} captures the exact critical point and phase boundaries separating the high and low current and high and low density regions. Attractive fixed points occur at $\alpha = \beta = 0.0$, the zero current fixed point, and at $\alpha = \beta \approx 2.929$, which attracts all points within the maximum current phase. The maximum current phase ($C$ in Fig.\ \ref{fig1}) is separated from the high and low density phases by second order phase boundaries, corresponding in the flow diagram to the two separatrices, each originating at the $\alpha = \beta = 0.5$ fixed point, with one attracted to the fixed point $(0.5, 2.929)$ and the other attracted to $(2.929,0.5)$. If one increases the length rescaling factor, these fixed points and the attractor for the maximum current phase should move toward their correct locations, i.e. the value $2.929$ should approach infinity.

An interesting closed subspace of the flow diagram is the line connecting $(0,1)$ and $(1,0)$, all contained within the low current region. On this line, $\alpha + \beta = 1$, the steady state solution becomes trivial. One can choose one dimensional matrices (scalars) $D=\beta^{-1}$ and $E=\alpha^{-1}$ to solve the problem. The flow diagram clearly captures this feature.

The basins of attraction corresponding to the high and low density regions are separated by a first-order boundary, evidenced in the flow diagram by the line from the unstable critical fixed point to the attractive fixed point at $\alpha = \beta = 0.0$. Thus, the flow diagram captures all of the phase boundaries and the critical point. A similar flow diagram was obtained by Stinchcombe et al. in \cite{RG_RDS4} by coarse graining the operators $D$ and $E$ and using them to calculate the system properties. Our results combined with previous treatments ~\cite{RG_RDS1, RG_RDS2, RG_RDS3, RG_RDS4} indicate that a reliable qualitative picture, and sometimes exact quantitative agreement, can be obtained with these position-space rescaling approaches using small length rescaling factors.

Central to the flow diagram is the critical point $\alpha_c = \beta_c = 0.5$, which is repulsive as expected.
The linearized recursion relations around this fixed point can be written as:
\begin{equation}
\left[\begin{array}{c}\delta\tilde\alpha\\
\delta\tilde\beta\end{array}\right] = \left[ \begin{array}{cc}\displaystyle {\frac{\partial\tilde\alpha}{\partial\alpha}\;\;
\frac{\partial\tilde\alpha}{\partial\beta} }\\
\displaystyle\frac{\partial\tilde\beta}{\partial\alpha}\;\; \displaystyle\frac{\partial\tilde\beta}{\partial\beta}\end{array}
\right]\!\!_{\begin{array}{c}\scriptstyle\alpha=1/2\\ \scriptstyle\beta=1/2 \end{array}}
\!\!\!\left[\begin{array}{c}\delta\alpha\\
\delta\beta\end{array}\right],\rm{\;\;\;where\;\delta\tilde\alpha=\tilde\alpha-1/2,\;etc.}
\label{EQ16}
\end{equation}

The above matrix has two eigenvalues $\lambda_1$ and $\lambda_2$.
From the ratio of the distances between consecutive
points in the renormalization-group flow (see Fig.\ \ref{fig4}), we obtain the numerical values for these eigenvalues
\begin{equation}
\lambda_1 = \lambda_2 = 1.5.
\label{EQ17}
\end{equation}
Thus the eigenvalue matrix is proportional to the identity matrix with a proportionality coefficient of $3/2$.
The critical exponent associated with the correlation length $\nu \equiv \ln b/\ln \lambda$ equals:
\begin{equation}
\nu=\frac{\ln (3)}{\ln (3/2)} \approx 2.710.
\label{EQ18}
\end{equation}

From the exact solution \cite{Schutz}, a length scale $\xi_\sigma$ can be defined:
\begin{equation}
\xi_\sigma = - \frac{1}{\ln[4\sigma (1-\sigma)]},
\label{EQ22}
\end{equation}
where $\sigma$ can be either $\alpha$ or $\beta$ and the length scale $\xi^{-1}\equiv{\xi_{\alpha}}^{-1} - {\xi_{\beta}}^{-1}$
governs the decay of the density profile. When $\sigma$ tends to $1/2$, this length scale diverges as:
\begin{equation}
\xi_\sigma \propto [\sigma - 1/2]^{-2},
\label{EQ23}
\end{equation}
which gives the critical exponent $\nu=2.00$. 

The same rescaling procedure can be applied to the more general system with probability $p\,dt$
for a jump to an empty site on the right. In this case Eqs.\ (\ref{EQ4}) become:
\begin{eqnarray}
\frac{d}{dt}\langle\tau_i\rangle &=& p\langle\tau_{i-1}(1-\tau_i)\rangle-p\langle\tau_{i}(1-\tau_{i+1})\rangle\nonumber\\
\frac{d}{dt}\langle\tau_1\rangle &=& \alpha\langle 1-\tau_1\rangle-p\langle\tau_1(1-\tau_2)\rangle\nonumber\\
\frac{d}{dt}\langle\tau_N\rangle &=& p\langle\tau_{N-1}(1-\tau_N)\rangle-\beta\langle\tau_N\rangle
\label{EQ19}
\end{eqnarray}
In terms of new variables $\hat{\alpha} = \alpha / p$ and $\hat{\beta} = \beta / p$, the steady
state Eqs.\ (\ref{EQ10}) become identical to the equations for the system with $p\,dt=1$. The critical point moves
to $\alpha_c=\beta_c=p/2$, in agreement with results obtained using other methods \cite{Sandow}, and the critical 
exponent stays the same.

\begin{figure}[htb]
\epsfig{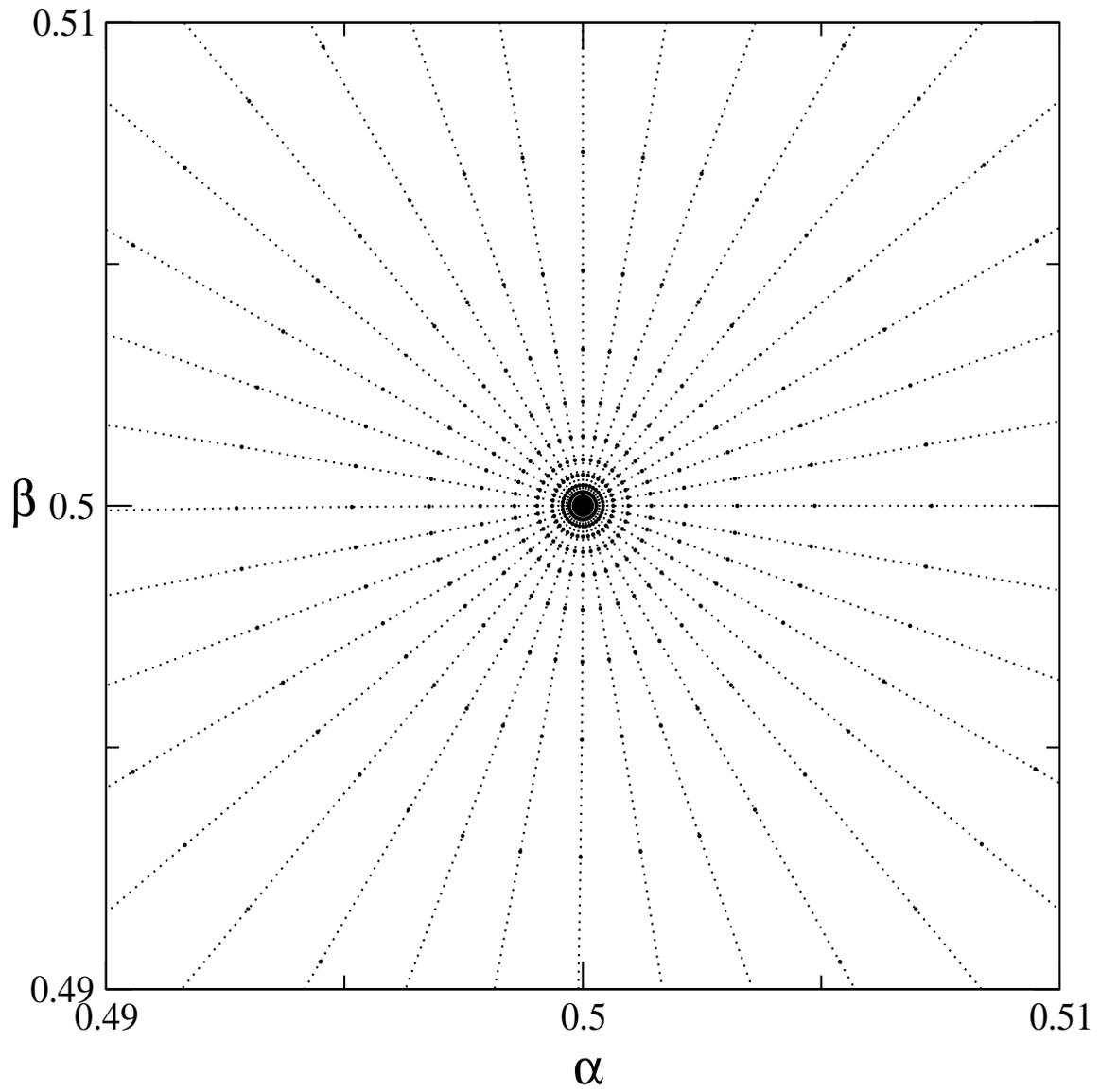}
\caption{Flow diagram in the vicinity of the critical point. Dots indicate the flow occurring at successive iterations away
from the unstable fixed point.}
\label{fig3}
\end{figure}

\begin{figure}[htb]
\epsfig{file=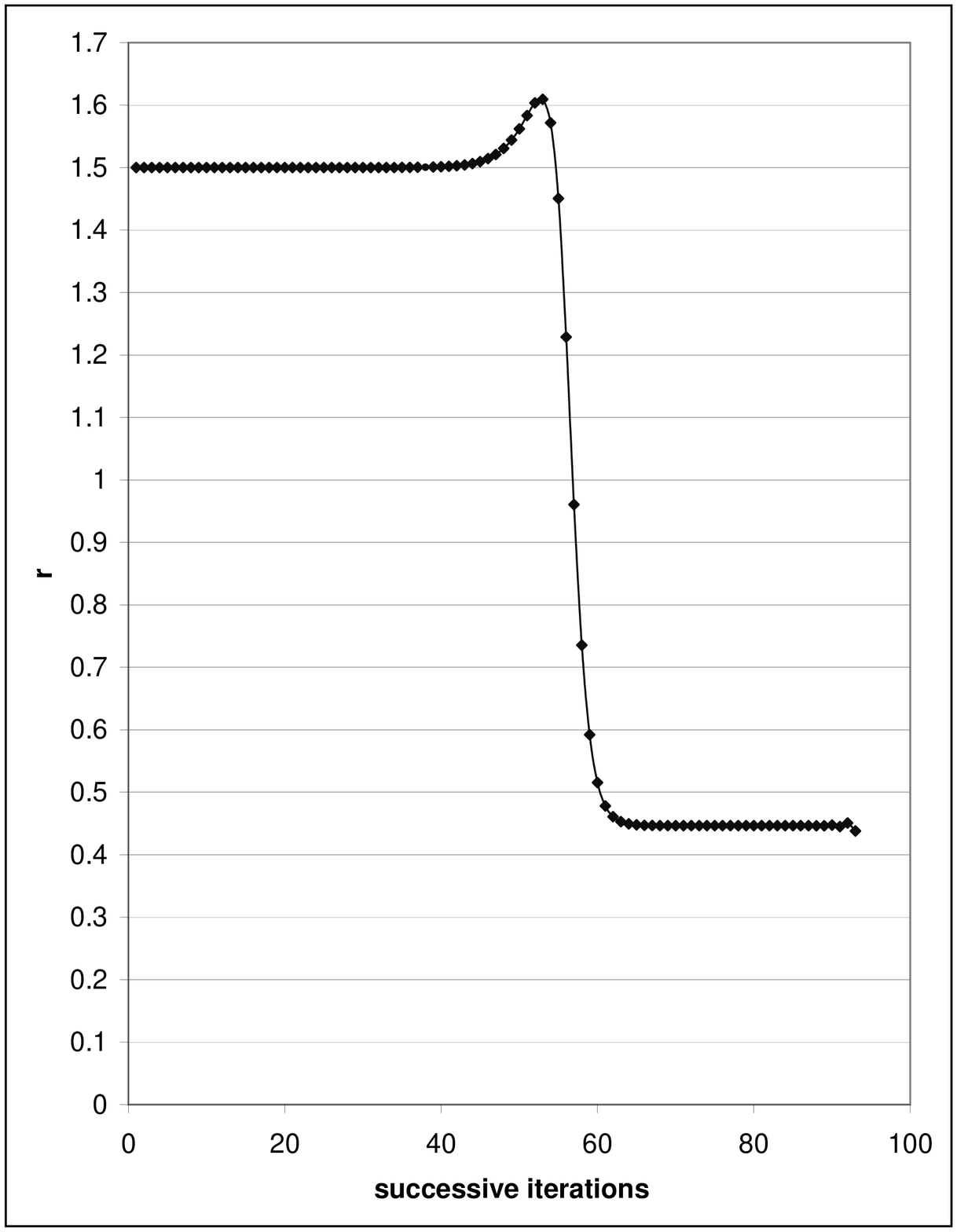, width=6in, height=6in}
\caption{The ratio $r$ of the length between successive points, which provides the eigenvalues of the matrix. The ratio of 
$1.5$ occurs in the critical region. The other ratio occurs near the attractor located at $(2.929, 2.929)$.}
\label{fig4}
\end{figure}

An interesting related question is whether the linearized recursion matrix Eq.\ (\ref{EQ16}) remains proportional to the 
identity matrix when larger rescaling factors are used (see Fig.\ ~\ref{fig3}). This conjecture can easily be proven. The general recursion relations 
between $(\tilde\alpha , \tilde\beta)$ and $(\alpha , \beta)$, because of
the particle-hole symmetry, would be of the form:
\begin{eqnarray}
\tilde\alpha=f[\alpha , J(\alpha, \beta)]\nonumber\\
\tilde\beta=f[\beta , J(\alpha, \beta)].
\label{EQ20}
\end{eqnarray}
The function $f(u,J)$ would be different for different scaling parameters (here $u$ can be either $\alpha$ or $\beta$).
It is easy to check that the matrix would become:
\begin{equation}
\left[ \begin{array}{cc}\displaystyle {\frac{\partial\tilde\alpha}{\partial\alpha}\;\;
\frac{\partial\tilde\alpha}{\partial\beta} }\\
\displaystyle\frac{\partial\tilde\beta}{\partial\alpha}\;\; \displaystyle\frac{\partial\tilde\beta}{\partial\beta}\end{array}
\right]\!\!_{\begin{array}{c}\scriptstyle\alpha=1/2\\ \scriptstyle\beta=1/2 \end{array}}
= \left[ \begin{array} {cc} \frac{\partial f}{\partial u}+\frac{\partial f}{\partial J} \frac{\partial J}{\partial\alpha} &  \frac{\partial f}{\partial J}\frac{\partial J}{\partial\beta}\\
\frac{\partial f}{\partial J}\frac{\partial J}{\partial\alpha} & \frac{\partial f}{\partial u}+\frac{\partial f}{\partial J} \frac{\partial J}{\partial\beta}
\end{array}\right]_{\alpha = \beta=1/2}
=\left[\frac{\partial f}{\partial u}\right]_{\alpha=\beta=1/2} \left[ \begin{array}{cc}1 & 0 \\ 0 &1\end{array}\right],
\label{EQ21}
\end{equation}
where Eqs. (\ref{EQ9}) are used to calculate the necessary derivatives in the different regions. Therefore the matrix remains
proportional to the identity matrix with proportionality coefficient $(\partial f/\partial u) $ evaluated at the critical point. The 
renormalization-group flow
does not distinguish between the high density regions $A_I$ and $A_{II}$ (or between the low density regions $B_I$ and
$B_{II}$) reported in \cite{Schutz}. These areas differ only in how the bulk density is approached, coming from the boundary 
site, and thus have identical macroscopic properties in the thermodynamic limit.

We have also tested another approach for constructing the recursion equations, one that imposes the requirement that the current remains invariant under rescaling, i.e. $\tilde J = J$. Applying this approach to the system with $p\,dt \neq 1$, again with a length 
rescaling factor of three, we obtain:
\begin{equation}
\tilde\alpha = \alpha \frac{\langle1-\tau_1 \rangle}{\langle 1-T_1 \rangle},\;\;
\tilde p = \alpha \frac{\langle1-\tau_1 \rangle}{\langle T_1(1-T_2)\rangle},\;\;
\tilde\beta = \beta \frac{\langle1-\tau_N \rangle}{\langle T_N\rangle},
\label{EQ24}
\end{equation}
and, with the matrix algebra changed to $pDE=D+E$, the recursion equations become:
\begin{eqnarray}
\tilde\alpha &=& \frac{1}{(\frac{1}{p^2}+\frac{1}{\alpha p}+\frac{1}{\alpha^2})J + \frac{1}{\alpha^2 p}J^2}\nonumber\\
\tilde\beta &=& \frac{1}{(\frac{1}{p^2}+\frac{1}{\beta p}+\frac{1}{\beta^2})J + \frac{1}{\beta^2 p}J^2}\\
\tilde p &=&\frac{1}{\frac{3}{p^2} J + \frac{4}{p^3} J^2 + (\frac{6}{p^4}+\frac{3}{\alpha p^3}+\frac{3}{\alpha^2 p^2})J^3 + (\frac{4}{p^5}-\frac{1}{\alpha p^4}+\frac{3}{\alpha^2 p^3}+\frac{3}{\alpha^3 p^2})J^4 + (\frac{4}{\alpha^2 p^4}+\frac{3}{\alpha^3 p^3})J^5 }\nonumber
\label{EQ25}
\end{eqnarray}

This approach yields the same value for the critical exponent $\nu=2.710$.
Here the parameters that change during the rescaling are: $\alpha, \beta$ and $p$.
As reported in \cite{Essler96}, the general case of the Fock representation of the
quadratic algebra involves twelve parameters that control the flow of the gas in the bulk of the chain. The general
steady state solution for this case is not known yet. As in the equilibrium case, in order to obtain
more accurate calculations, we would have to include in the system after rescaling new dynamical rules, to
add more allowed transitions between states. In other words, the rescaled dynamics, with appropriate generality, 
should include possibilities for the following transitions:
\begin{eqnarray}
\text{Diffusion to the right:}  \;\;\;\;&1& + \;0 \rightarrow 0 + 1, \;\;\text{( rate $\Gamma_{01}^{10}$)}\nonumber\\
\text{Coagulation at the right:}  \;\;\;\;&1& + \;1 \rightarrow 0 + 1, \;\;\text{( rate $\Gamma_{01}^{11}$)}\nonumber\\
\text{Decoagulation at the right:}  \;\;\;\;&1& + \;0 \rightarrow 1 + 1, \;\;\text{( rate $\Gamma_{11}^{10}$)}\nonumber\\
\text{Birth at the right:}  \;\;\;\;&0& + \;0 \rightarrow 0 + 1, \;\;\text{( rate $\Gamma_{01}^{00}$)}\nonumber\\
\text{Death at the right:}  \;\;\;\;&1& + \;0 \rightarrow 0 + 0, \;\;\text{( rate $\Gamma_{00}^{10}$)}
\label{EQ26}
\end{eqnarray}

\section{CONCLUSIONS}
We have presented a general position-space renormalization-group approach for driven diffusive systems and shown how it 
can be applied to the asymmetric exclusion model.  The same scheme can be applied to any system in which the parameters 
driving the system can be expressed in terms of the system's correlation functions, provided that these correlation functions 
can be conveniently stated.  Thus, the crucial part is that, when the system is rescaled,
the resulting higher correlations which enter into the equations can be calculated exactly or within a good approximation. 
In the cases discussed above, these correlations
are reducible to a functional dependence on the steady state current.

The method that we have introduced to study the totally asymmetric case is not only
interesting in itself, but also shows how the rescaling procedure can be applied to systems out of equilibrium
to determine critical properties when the steady state is known exactly or to a good approximation.
The second method for deriving the recursion relation, i.e. $J = \tilde J$,
provides a general scheme to study other systems such as models of fast ionic conductors, gel electrophoresis, traffic flows, etc.,
within a position-space renormalization-group framework.

We thank Beate Schmittmann and Royce Zia for helpful discussions.  This material is based upon work supported in part by the National Science Foundation under Grant No.  9720482.

\section{APPENDIX}

Here we show the details for obtaining  Eq. (\ref{EQ15}). Using the algebraic rules:

\begin{eqnarray}
C & \equiv &D+E\nonumber\\ 
DE &=& D+E\nonumber\\
D\vert V\rangle &=& \frac{1}{\beta}\vert V\rangle\nonumber\\
\langle W\vert E &=& \frac{1}{\alpha}\langle W\vert .
\label{AEQ1}
\end{eqnarray}

and the obvious consequences of them:

\begin{eqnarray}
D &=& C-E\nonumber\\
D^2 &=& C^2 - EC - C\nonumber\\
D^3 &=& C^3 - 2C^2 - EC^2 + EC ,\;\;\;etc ... .
\label{AEQ2}
\end{eqnarray}
one can calculate the expressions:
\begin{eqnarray}
\langle DC^{N}\rangle &=& \langle C^{N+1}\rangle - \frac{1}{\alpha}\langle C^N\rangle\nonumber\\
\langle D^2C^{N}\rangle &=& \langle C^{N+2}\rangle - (1+\frac{1}{\alpha})\langle C^{N+1}\rangle\nonumber\\\
\langle D^3C^{N}\rangle &=& \langle C^{N+3}\rangle - (2+\frac{1}{\alpha})\langle C^{N+2}\rangle + \frac{1}{\alpha}\langle C^{N+1}\rangle
\label{AEQ3}
\end{eqnarray}
In an analogous way, one can derive the formulas involving $E$ and $\beta$. For example, below we show how the calculation 
for the expression in the
denominator in  Eq.(\ref{EQ11}) is done.
\begin{eqnarray}
\langle 1-T_1\rangle_T &=& \langle100\rangle_\tau + \langle010\rangle_\tau + \langle001\rangle_\tau+ \langle000\rangle_\tau\nonumber\\
\langle100\rangle_\tau &=& \frac{\langle W|DE^2C^{N-3}|V\rangle_\tau}{\langle W|C^N|V\rangle_\tau}\nonumber\\
\langle010\rangle_\tau &=& \frac{\langle W|EDEC^{N-3}|V\rangle_\tau}{\langle W|C^N|V\rangle_\tau}\nonumber\\
\langle001\rangle_\tau &=& \frac{\langle W|E^2DC^{N-3}|V\rangle_\tau}{\langle W|C^N|V\rangle_\tau}\nonumber\\
\langle000\rangle_\tau &=& \frac{\langle W|E^3C^{N-3}|V\rangle_\tau}{\langle W|C^N|V\rangle_\tau}
\label{AEQ4}
\end{eqnarray}
In order to calculate the average in the numerator for $\langle 100\rangle_\tau$ we rewrite $ DE^2C^{N-3}$ as:
\begin{equation}
DE^2C^{N-3} = CEC^{N-3}  = (C+E^2)C^{N-3}= C^{N-2}  - E^2 C^{N-3}
\label{AEQ5}
\end{equation}
Now it can easily be calculated using Eqns. (\ref{AEQ3}) to give:
\begin{equation}
\frac{\langle W|DE^2C^{N-3}|V\rangle_\tau}{\langle W|C^N|V\rangle_\tau} = \frac{\langle W|C^{N-2}|V\rangle_\tau}{\langle W|C^N|V\rangle_\tau} + \frac{\langle W|E^2C^{N-3}|V\rangle_\tau}{\langle W|C^N|V\rangle_\tau} = J^2 + \frac{1}{\alpha^2}J^3
\label{AEQ6}
\end{equation}
In the same way we obtain the rest of the averages:
\begin{eqnarray}
\langle010\rangle_\tau &=& \frac{\langle W|EC^{N-2}|V\rangle_\tau}{\langle W|C^N|V\rangle_\tau} = \frac{1}{\alpha}J^2\nonumber\\
\langle001\rangle_\tau &=& \frac{\langle W|E^2DC^{N-3}|V\rangle_\tau}{\langle W|C^N|V\rangle_\tau}= \frac{1}{\alpha^2}J^2 - \frac{1}{\alpha^3}J^3\nonumber\\
\langle000\rangle_\tau &=& \frac{\langle W|E^3C^{N-3}|V\rangle_\tau}{\langle W|C^N|V\rangle_\tau} =\frac{1}{\alpha^3}J^3
\label{AEQ7}
\end{eqnarray}
Combining these expressions leads to the result for $\langle 1-T_1\rangle_T = (1+\frac{1}{\alpha}+\frac{1}{\alpha^2}+\frac{J}{\alpha^2})J^2$ .
The rest of the calculations are done using the same techniques. The case of $p \neq 1$ can be handled in the same manner.

\pagebreak

\pagebreak

\pagebreak

\pagebreak


\begin{references}
\bibitem{ziabook} B. Schmittmann and R. K. P. Zia, {\it Statistical Mechanics of Drive Diffusive Systems}, (Academic Press 1995).
\bibitem{Der93} B. Derrida, M.R. Evans, V. Hakim and V. Pasqueier, J. Phys. A: Math Gen. {\bf 26}, 1493 (1993).
\bibitem{Der_book} B. Derrida and M. R. Evans, in: {\it Nonequilibrium Statistical Mechanics in One Dimension} ed. V. Privman, (Cambridge University
Press 1997).
\bibitem{Krug} J. Krug, Phys. Rev. Lett. {\bf 67}, 1882 (1991).
\bibitem{ren1} R.J. Creswick, H.A. Farach, C.P. Poole, Jr., {\it Intoduction to Renormalization Group Methods in Physics}, (John Wiley \& Sons, Inc. 1992).
\bibitem{ren2} Annick Lesne, {\it Renormalization Methods}, (John Wiley \& Sons, Inc. 1998).
\bibitem{Der92} B. Derrida, E. Domany, and D. Mukamel, J. Stat. Phys. {\bf 69}, 667 (1992).
\bibitem{Schutz} G. Schutz and E. Domany, J. Stat. Phys. {\bf 72}, 277 (1993).
\bibitem{Sandow} S. Sandow, Phys. Rev. E {\bf 50}, 2660 (1994).
\bibitem{Essler96} F. H. L. Essler and V. Rittenberg, J. Phys. A: Math. Gen. {\bf 29}, 3375 (1996).
\bibitem{RG_RDS1} J. Hooyberghs and C. Vanderzande, J. Phys. A: Math. Gen. {\bf 33}, 907 (2000).
\bibitem{RG_RDS2} J. Hooyberghs, E. Carlon, and C. Vanderzande, Phys. Rev. E. {\bf 64}, 036124 (2001).
\bibitem{RG_RDS3} J. Hooyberghs and C. Vanderzande, Phys. Rev. E. {\bf 63}, 041109 (2001).
\bibitem{RG_RDS4} R. B. Stinchcombe and T. Hanney, Physica D {\bf 168}, 313 (2002).
\end{references}
\end{document}